\documentclass[12pt]{article}

\usepackage{amssymb}

\usepackage{cite}
\usepackage{graphicx}
\def\labelmark{}
\def\void{}

{\ifx\void\labelname\def\junk{\end{displaymath}}
\else\def\junk{\end{eqnarray}}\fi\junk\labelmark\def\labelname{}}

\newcommand{\bra}{\begin{array}}
\newcommand{\era}{\end{array}}
\newcommand{\beq}{\begin{equation}}
\newcommand{\eeq}{\end{equation}}
\newcommand{\bqn}{\begin{eqnarray}}
\newcommand{\eqn}{\end{eqnarray}}

 \font\mybb=msbm10  at 12pt
\def\bb#1{\hbox{\mybb#1}}

\font\mybbi=msbm10  at 9pt
\def\bbi#1{\hbox{\mybbi#1}}

\def\BC{\bb C}
\def\_\BC{\bbi C}
\def\RR{\bb R}

\newcommand{\om}{\omega}

\newcommand{\eps}{\epsilon}

\newcommand{\te}{\theta}

\newcommand{\pa}{\partial}
\newcommand{\al}{\alpha}

\newcommand{\del}{\delta}
\newcommand{\lga}{\longrightarrow}

\newcommand{\st}{\star}

\newcommand{\da}{\dagger}

\newcommand{\ov}{\over}
\newcommand{\hb}{\hbar}

\newcommand{\ev}{\equiv}
\newcommand{\lb}{\label}

\begin{document}
\begin{titlepage}
\setcounter{page}{1}
\renewcommand{\thefootnote}{\fnsymbol{footnote}}

\begin{flushright}
hep-th/0309105
\end{flushright}

\vspace{6mm}
\begin{center}

{\Large\bf Two Coupled Harmonic Oscillators \\ on Non-commutative
Plane}

\vspace{8mm}
{\bf Ahmed Jellal$^1$\footnote{E-mail:
{\textsf jellal@gursey.gov.tr }}},
{\bf El Hassan El Kinani$^{2,3}$\footnote{E-mail:
{\textsf hkinani@ictp.trieste.it }}}
 and {\bf Michael Schreiber$^1$}

\vspace{5mm} $^1${\em  Institut f\"{u}r Physik, Technische
Universit\"{a}t, \\ D-09107 Chemnitz, Germany }\\

$^2$ {\em The Abdus Salam International Center for Theoretical Physics,\\
 Strada Costiera 11, 34014 Trieste, Italy}\\

$^3$ {\em  Moulay Ismail University, Faculty of Science and Technical\\
Mathematics Department, P.O. Box 509, Boutalamine, Errachidia, Morocco}

\vspace{5mm}
\begin{abstract}

We investigate a system of two coupled harmonic oscillators on
the non-commutative plane $\RR^2_{\te}$ 
by requiring that the spatial coordinates do not commute. 
We show that the system can be
diagonalized by a suitable transformation,
i.e. a rotation with a mixing angle $\al$. The obtained
eigenstates as well as the
eigenvalues depend on the non-commutativity parameter
$\te$. Focusing on the ground state wave function before 
the transformation, we calculate the density matrix
$\rho_0(\te)$ 
and find that its traces
${\rm Tr}\left(\rho_{0}(\te)\right)$ and 
${\rm Tr}\left(\rho_0^2(\te)\right)$ 
are not affected by
the non-commutativity. 
Evaluating the Wigner
function on $\RR^2_{\te}$ confirms this. 
The uncertainty relation is
explicitly determined and 
found to depend on $\te$.  
For small values of $\te$,
the relation is shifted by a $\te^2$ term, 
which can be interpreted as a quantum correction. 
The calculated entropy does not change with respect
to the normal case. 
We consider the limits $\al=0$  
and $\al={\pi\ov 2}$. In first case, by identifying
$\te$ to the squared magnetic length,
one can recover basic features of the Hall system.

\end{abstract}
\end{center}
\end{titlepage}

\section{Introduction}

Because of its mathematical simplicity, the harmonic oscillator
provides solvable models in many branches of physics. It often
gives a clear illustration of an abstract idea. For instance a
charged particle on the plane ${\vec x} = (x_1,x_2)$ 
in presence of a strong uniform magnetic field
described by the Lagrangian
\begin{equation}
L = {m\ov 2} \dot{\vec r}^2 + {\vec A}\cdot \dot{\vec x} -V({\vec x})
\end{equation}
where the vector potential ${\vec A}=(A_1,A_2)$ 
and the confining potential are
\begin{equation}
\lb{pot}
A_i={1\ov 2}B\eps_{ij}x_j, \qquad V({\vec x})= {1\ov 2}\om {\vec x}^2
\end{equation}
with $\eps_{12}= -\eps_{21}=1$ and $0$ otherwise,
gives a nice natural non-commutative
system~\cite{dune} and also a good starting point to discuss
the quantum Hall effect~\cite{prange}. Now it is natural 
to ask what happens if the external potential is not
parabolic? 
This question was answered by
Kim {\it et al.} $[3-11]$ starting more than 
twenty years ago. They considered 
two coupled harmonic oscillators with the potential 
\begin{equation}\label{kpot}
V({\vec x}) = {1\over
2}\left( c_1 x_{1}^{2} + c_2 x^{2}_{2} + c_3 x_{1} x_{2} \right)
\end{equation}
with constant $c_1, c_2, c_3$ and explicitly determined
the corresponding density matrix and the Wigner function 
as well as other quantities.

There are many physical models based on coupled harmonic
oscillators, such as the Lee model in quantum field
theory~\cite{sss61}, the Bogoliubov transformation in
superconductivity~\cite{fewa71}, two-mode squeezed states of
light~\cite{hkn90,dir63,cav85}, the covariant harmonic oscillator
model for the parton picture~\cite{kim89}, and models in molecular
physics~\cite{iac91}.  There are also models of current interest
in which one of the variables is not observed, including
thermo-field dynamics~\cite{ume82}, two-mode squeezed
states~\cite{yupo87,ekn89}, the hadronic temperature~\cite{hkn89},
and the Barnet-Phoenix version of information
theory~\cite{baph91}.  In all of these cases, 
the mixing angle 
$\alpha$ of the employed transformation 
is ${\pi\ov 2}$, and in this situation
the mathematics becomes simple.

In this paper we study quantum mechanically a system of
two coupled harmonic oscillators  
on the non-commutative (NC) plane $\RR^2_{\te}$.
This can be done by demanding that the spatial coordinates
do not commute.
We use the star-product to write the NC
Hamiltonian and solve the eigenequations to get the
eigenstates as well as the energy spectrum.
Focusing on the ground state, we evaluate the
corresponding density matrix. It depends on the
non-commutativity parameter $\theta$ but
its traces as well as the entropy are not affected
by the non-commutativity.
Also we calculate the Wigner function to confirm 
the  $\theta$-independence of the traces.
The uncertainty relation on $\RR^2_{\te}$
is found to depend on $\theta$ and it coincides with
the normal case in the limit $\theta=0$.
For small $\theta$, we show that
this relation contains a quantum correction,
which is a shift with a $\theta^2$ term.
Also we discuss some limits of
the mixing angle, namely $\al={\pi\ov 2}$ and $\al=0$. In 
the last case, 
the system can be linked to the Hall electron.

In section 2, we give the energy spectrum and the eigenstates of a
Hamiltonian describing two coupled harmonic oscillators. This
serves as a guide in section 3 where we consider the same system but
require that the spatial coordinates do not commute.
In section 4, we deal with the corresponding density matrix and
evaluate its traces. In section 5 we calculate the
Wigner function on $\RR^2_{\te}$ and investigate 
its link with the density matrix. We determine explicitly the
uncertainty relation as well as the entropy
in section 6. Some particular cases will be considered
in section 7. Finally, we conclude our work in the last section.

\section{Coupled harmonic oscillators}

Let us consider a system of two coupled harmonic oscillators
parameterized by the coordinates $X_1, X_2$ and masses $m_1,
m_2$. This can be described by a Hamiltonian as the sum of free
and interacting parts \cite{hkn99am}
\begin{equation}\label{HAM1}
H_1 = {1\over 2m_{1}}P^{2}_{1} + {1\over 2 m_{2}} P^{2}_{2} +
{1\over 2} \left( C_1X^{2}_{1} + C_2 X^{2}_{2} + C_3 X_{1}
X_{2}\right)
\end{equation}
where $C_1, C_2, C_3$ are constant parameters. 
After rescaling the position variables
\begin{equation}
x_{1} = ({m_{1}/ m_{2}})^{1\ov 4} X_{1},
\qquad
x_{2} = ({m_{2}/ m_{1}})^{1\ov 4} X_{2}
\end{equation}
as well as the momentum
\begin{equation}
p_{1} = ({m_{2}/ m_{1}})^{1\ov 4} P_{1},
\qquad
p_{2} = ({m_{1}/m_{2}})^{1\ov 4} P_{2}
\end{equation}
$H_1$ can be written as
\begin{equation}\label{HAM2}
H_2 = {1\over 2m}\left(p^{2}_{1} + p^{2}_{2} \right) + {1\over
2}\left( c_1 x_{1}^{2} + c_2 x^{2}_{2} + c_3 x_{1} x_{2} \right)
\end{equation}
where the parameters are
\begin{equation}\label{para}
m = (m_{1}m_{2})^{1/2},\qquad c_1=C_1\sqrt{m_2\over m_1}, \qquad
c_1=C_1\sqrt{m_1\over m_2},\qquad c_3=C_3.
\end{equation}

As the Hamiltonian~(\ref{HAM2}) involves an interacting
term, a straightforward investigation of the basic features of the
system is not easy. Nevertheless, we can simplify this
situation by a transformation to new phase space variables 
\begin{equation} 
\lb{TRAN} y_a = M_{ab} x_b, \qquad  q_a = M_{ab} p_b 
\end{equation}
where the matrix 
\begin{equation} \label{UMAT}
(M_{ab}) = \pmatrix{\cos {\alpha\ov 2} & -\sin {\alpha\ov 2} \cr
\sin {\alpha\ov 2} & \cos {\alpha\ov 2}}
\end{equation}
is a unitary rotation with the mixing angle
$\al$. Inserting the mapping~(\ref{TRAN}) into~(\ref{HAM2}), one
realizes that $\al$ should satisfy the condition
\begin{equation}\label{acon}
\tan \alpha  = {c_3\over c_2 - c_1}
\end{equation}
to get a factorizing Hamiltonian
\begin{equation}
\label{HAM3} H_3 = {1\over 2m} \left(q^{2}_{1} + q^{2}_{2} \right) +
{K\over 2}\left(e^{2\eta } y^{2}_{1} + e^{-2\eta }
y^{2}_{2}\right)
\end{equation}
where  
\begin{equation}\label{PARA}
K = \sqrt{c_1c_2 - c_3^{2}/4} , \qquad  e^{\eta}= \frac {c_1 + c_2
+ \sqrt{(c_1 - c_2)^{2} + c_3^{2}}}{2K}
\end{equation}
and the condition $4c_1c_2 > c_3^{2}$ must be fulfilled.

It is convenient to
separate~(\ref{HAM3}) into two commuting parts
\begin{equation}
\label{HAM4} H_3 = e^{\eta} {\cal H}_1 + e^{-\eta} {\cal H}_2
\end{equation}
where ${\cal H}_1$ and ${\cal H}_2$ are given by
\begin{equation}
\label{HAM41} {\cal H}_1 = {1\ov 2m}e^{-\eta}q^{2}_{1} +
{K\ov 2} e^{\eta}y^{2}_{1} , \qquad 
{\cal H}_2 = {1\ov 2m}e^{\eta}q^{2}_{2} + 
{K\ov 2} e^{-\eta}y^{2}_{2}.
\end{equation}
First, one can see that the decoupled Hamiltonian 
\begin{equation}
\label{HAM5} H_0 = {1\ov 2m}q^{2}_{1} +
{K\ov 2} y^{2}_{1} + {1\ov 2m}q^{2}_{2} +
{K\ov 2}y^{2}_{2}
\end{equation}
is obtained for $\eta=0$.
Second, it is interesting to note that~(\ref{HAM5}) can be
derived by a canonical transformation only from
\begin{equation}
\label{HAM6} {\cal H} = {\cal H}_1 + {\cal H}_2
\end{equation}
rather than from~(\ref{HAM4}). This suggests that it might be 
advantageous
to consider (\ref{HAM6}) instead of~(\ref{HAM4}). Because
of that Kim {\it et al.}~\cite{hkn95jm,hkn99am} were focusing
on the Hamiltonian~(\ref{HAM6}).

It is clear that ${\cal H}$ is a Hamiltonian of two decoupled
harmonic oscillators. Thus it can simply be diagonalized
by defining a set of creation and annihilation operators
\begin{equation}
\lb{CRAN} 
a_i = \sqrt{K\ov 2\hb\om} e^{\eta\ov 2}y_{i} +
{i \ov \sqrt{2m\hb\om}} e^{-{\eta\ov 2}}q_{i}, \qquad a_i^{\da} =
\sqrt{K\ov 2\hb\om} e^{\eta\ov 2}y_{i} - {i \ov \sqrt{2m\hb\om}}
e^{-{\eta\ov 2}}q_{i} 
\end{equation}
with frequency 
\begin{equation}
\om=\sqrt{K\ov m}.
\end{equation}
They satisfy the commutation relations
\begin{equation}
 [a_i, a_j^{\dag}] = \delta_{ij}.
\end{equation}
Other commutators vanish. Now we can map ${\cal H}$ in terms of
$a_i$ and $a_i^{\dag}$ as
\begin{equation}
\label{HAM7} {\cal H} = \hb\om \left(a_1^{\dag}a_1 + a_2^{\dag}a_2 +
1\right).
\end{equation}

To obtain the eigenstates and the eigenvalues, one
solves the eigenequations
\begin{equation}
 {\cal H}|n_1, n_2\rangle = {\cal E}_{n_1,n_2} |n_1,
n_2\rangle
\end{equation}
getting
\begin{equation}
|n_1,n_2\rangle= {(a_1^{\dag})^{n_1} (a_2^{\dag})^{n_2} \ov
\sqrt{n_1!n_2!}} |0, 0\rangle
\end{equation}
where the ground state wave function is
\begin{equation}
\label{YWAV0}
\psi_0(\vec{y})\ev \langle y_1,y_2|0,0\rangle
= \sqrt{m\om \ov \pi\hb}
\exp{\left\{-{m\om \over 2\hb}\left(e^{\eta} y^{2}_{1} + e^{-\eta}
y^{2}_{2}\right) \right\} }
\end{equation}
as well as the energy spectrum
\begin{equation}
\label{SPE1} {\cal E}_{n_1,n_2} = \hb\om \left( n_1 + n_2+ 1\right).
\end{equation}

The above solutions can be used to deduce those corresponding to
$H_3$, in particular its energy spectrum
\begin{equation}
\label{SPE2} E_{3,n_1,n_2} = {\hb\om} \left(e^{\eta} \left(n_1
+{1\ov 2}\right) +e^{-\eta} \left(n_2 +{1\ov 2}\right)\right)
\end{equation}
and the ground state wave function
\begin{eqnarray}
\lefteqn{
\lb{XWAV} \psi_{0}(\vec{x}) \ev
\langle x_1,x_2|0,0\rangle} 
\nonumber \\
& &
~~
= \sqrt{m\om \ov \pi\hb}
\exp\left\{-{m\om \over 2\hb}\left[e^{\eta}\left(x_{1}
\cos{\alpha\over 2} - x_{2} \sin{\alpha\over 2}\right)^{2} +
e^{-\eta}\left(x_{1}\sin{\alpha\over 2} + x_{2}\cos{\alpha\over
2}\right)^{2} \right] \right\}.
\end{eqnarray}
While~(\ref{YWAV0}) is separable in terms of the variables
$y_{1}$ and $y_{2}$, this is not the case
for~(\ref{XWAV}) in terms of $x_{1}$ and $x_{2}$.

In what follows, we generalize the present system to
the NC case by deforming the spatial configuration. This
will be used to investigate some physical quantities of the
system, i.e. the density matrix and the Wigner function as well
as other quantities corresponding to the ground state wave 
function (\ref{XWAV}) in the NC case.

\section{Non-commutative system}

We proceed by using the NC
geometry~\cite{connes} to study two coupled harmonic oscillators.
We demand that the coordinates of the plane do not commute
\begin{equation}
[x_{i},x_{j}]=i\te_{ij} \label{nccoo}
\end{equation} where
$\te_{ij}=\eps_{ij}\te$ is the non-commutativity parameter. 
This relation can be obtained using
the star-product definition
\begin{equation}
\lb{defi} f(x) \st g(x)=\exp\left\{{i\over 2}\te_{ij}
\pa_{x^{i}}\pa_{y^{j}}\right\} f(x)g(y){\Big{|}}_{x=y} \label{2}
\end{equation}
where $f$ and $g$ are two arbitrary
functions, supposed to be infinitely differentiable. In what
follows, we will use the standard commutation relations
\begin{equation}
\lb{deqm}
[p_{i},x_{j}]=-i\del_{ij}, \qquad [p_{i},p_{j}]=0 
\end{equation}
supplemented by the relation~(\ref{nccoo}). Together they define a
generalized quantum mechanics, which leads to the standard one for
$\te=0$.

Now let us define the Hamiltonian~(\ref{HAM2}) on $\RR_{\te}^2$.
Noting that $H_2$ acts on an arbitrary function
$\Psi(\vec{r},t)$ as
\begin{equation}
\lb{hdef} 
H_2 \st \Psi (\vec{r},t) = H_2^{\rm nc}
\Psi (\vec{r},t)
\end{equation}
and applying the definition~(\ref{defi}) we obtain
\begin{equation}\label{NHA0}
H_2^{\rm nc} = {1\over 2m}\left(p^{2}_{1} + p^{2}_{2}
\right) + {c_1\over 2} \left(x_{1}-{\te\ov 2\hb}p_2 \right)^{2} +
{c_2\over 2} \left(x_{2}+{\te\ov 2\hb}p_1 \right)^{2}
+ {c_3\over 2} \left(x_{1}-{\te\ov 2\hb}p_2\right )
\left(x_{2}+{\te\ov 2\hb}p_1\right ). 
\end{equation}
With~(\ref{NHA0}), we actually have two possibilities to get the
NC version of~(\ref{HAM6}). This can be done either by
transforming ${H_2^{\rm nc}}$ via~(\ref{TRAN}) to obtain
\begin{equation}
 \lb{NHA1} {\cal H}^{\rm nc} ={1\ov 2M}\left(e^{-\eta}{q}_{1}^2 +
e^{\eta}{q}_{2}^2\right) + {K\ov 2} \left(e^{\eta} y_1^2 +
e^{-\eta} y_2^2 \right)+ {K\te\ov 2\hb} \left(e^{-\eta}y_2q_1 -
e^{\eta}y_1q_2\right)
\end{equation}
or by starting straightforwardly from~(\ref{HAM6}), using
(\ref{defi}) to end up with~(\ref{NHA1}). The effective mass is 
given by
\begin{equation}
M = {m\ov 1 +\left({m\om \te\ov 2\hb}\right)^2}.
\end{equation}
It is useful to write~(\ref{NHA1}) in the compact form
\begin{equation}
\lb{NHA2} {\cal H}^{'\rm nc} = {1\ov 2M}\left({Q}_{1}^2 +
{Q}_{2}^2\right) + {K\ov 2} \left( Y_1^2 + Y_2^2 \right)+ {K\te\ov
2\hb} \left(Y_2Q_1 - Y_1Q_2\right)
\end{equation}
 by rescaling the
variables to new coordinates $Y_i$ and $Q_i$. Comparing ${\cal H}^{'\rm
nc}$ to ${\cal H}$, we note that ${\cal H}^{'\rm nc}$ contains an additional
term proportional to $\te$ which is basically the
angular momentum. Also, the NC system shows an effective mass $M$
which coincides with the mass $m$ for $\te=0$.

For the diagonalization of ${\cal H}^{'\rm nc}$ we
express the position and momentum variables in terms of
creation and annihilation operators
\begin{equation} 
Y_i =\sqrt{\hb\Omega\ov 2K} \left(b_i+b_i^{\dag} \right),\qquad Q_i = i
\sqrt{M\hb\Omega\ov 2} \left(b_i^{\dag} -b_i\right) 
\end{equation}
 which commute and satisfy the relations
\begin{equation}
 \lb{bCOM}
[b_i, b_j^{\dag}] = \del_{ij}
\end{equation}
where the effective frequency 
\begin{equation}
\Omega = \sqrt{K\ov M}
\end{equation}
depends on $\theta$.  With the help of 
another set of operators
\begin{eqnarray}
\lb{BOP} 
\lefteqn{
~~B_1 = {1\ov\sqrt{2}} (b_1+ib_2),\qquad 
B_1^{\dag}= {1\ov\sqrt{2}} (b_1^{\dag}-ib_2^{\dag})}
\nonumber \\
&
B_2 = {1\ov\sqrt{2}} (-b_1+ib_2),\qquad B_2^{\dag} =
{1\ov\sqrt{2}} (-b_1^{\dag}-ib_2^{\dag}) 
\end{eqnarray}
which satisfy
\begin{equation}
 \lb{BCOM} [B_i, B_j^{\dag}] = \del_{ij},
\end{equation}
one can write
\begin{equation}
\lb{NHA3} 
{\cal H}^{'{\rm nc}} = \hb\Omega_1 B_1^{\dag}B_1 +
\hb\Omega_2 B_2^{\dag}B_2 + \hb\Omega 
\end{equation}
with frequencies
\begin{equation}
\Omega_1 = \Omega +{K\te\ov 2\hb},
\qquad
\Omega_2 = \Omega - {K\te\ov 2\hb}.
\end{equation}

With~(\ref{NHA3}), we can easily solve the eigenequations
\begin{equation}
{\cal H}^{'\rm nc} |n_1, n_2,\te \rangle = {\cal E}^{'\rm
nc}_{n_1, n_2}|n_1, n_2,\te \rangle
\end{equation}
and obtain
\begin{equation}
 |n_1, n_2,\te \rangle= {(B_1^{\dag})^{n_1}
(B_2^{\dag})^{n_2} \ov \sqrt{n_1!n_2!}} |0, 0,\te \rangle
\end{equation}
 and the
eigenvalues
\begin{equation}
\label{NSPE1} {\cal E}_{n_1,n_2}^{'\rm nc} = \hb\Omega_1 n_1
+\hb\Omega_2 n_2+ \hb\Omega.
\end{equation}
Projecting $|0, 0,\te \rangle$
on the plane $(Y_1,Y_2)$ we find 
the ground state wave function
\begin{equation}
\label{YWAV} \psi_0(\vec{Y},\te) = \sqrt{M\Omega \ov \pi\hb}
\exp{\left\{-{M\Omega \over 2\hb}\left(Y^{2}_{1} +
Y^{2}_{2}\right) \right\} }.
\end{equation}
In terms of the $\vec{x}$ representation, it can be written as
\begin{equation}
\lb{NXWAV} \psi_{0}(\vec{x},\theta) = \sqrt{M\Omega \ov \pi\hb}
\exp\left\{-{M\Omega \over 2\hb}\left[e^{\eta}\left(x_{1}
\cos{\alpha\over 2} - x_{2} \sin{\alpha\over 2}\right)^{2} +
e^{-\eta}\left(x_{1}\sin{\alpha\over 2} + x_{2}\cos{\alpha\over
2}\right)^{2} \right] \right\}.
\end{equation}
We note that
the analysis in the previous section is recovered 
for $\te=0$.

\section{Density matrix of the NC system}

Because of its relevance in determining several thermodynamic
quantities, it is interesting to calculate the 
density matrix 
\begin{equation}
\lb{RDEF} \rho(\vec{x},\vec{x}') = \psi(\vec{x})\psi^*(\vec{x}')
\end{equation}
of the NC system. 

For the ground state wave function~(\ref{NXWAV})
the density matrix 
\begin{equation}
\rho_{0}(\vec{x},\vec{x}',\te) =
\psi_0(\vec{x},\te)\psi_0^*(\vec{x}',\te)
\end{equation}
can be represented as an integral
\begin{equation}
\rho_{0}(\vec{x},\vec{x}',\te)=
\int \rho_0(\vec{x},\vec{x}'',\te)
\rho_0(\vec{x}'',\vec{x}',\te) \; dx_{1}'' dx_{2}''.
\end{equation}
Tracing $\rho_0(\vec{x},\vec{x}',\te)$ over the variable
$x_{2}$, the resulting density is
\begin{equation}\label{integ}
\rho_0(x_{1},x_{1}',\te) = \int \psi_0(x_{1},x_{2},\te)
\psi_0^{*}(x_{1}',x_{2},\te) \; dx_{2} .
\end{equation}
Evaluating this integral, we get
\begin{eqnarray}
\label{NCDM} 
\lefteqn{
\rho_0(x_{1},x_{1}',\te) = \left({\Omega
M\over \pi \hb \gamma}\right)^{1/2} 
\exp\left\{{\Omega M\over 4\hb\gamma}
\left(x_{1} + x_{1}'\right)^{2} 
{\sin^2 \al \sinh^2\eta}\right\}~{} }
\nonumber \\
& & \qquad~~~~~~~~\exp\left\{-{\Omega M\over 2\hb}
\left(x_{1}^2 + x_{1}^{'2}\right)
\left(e^{\eta}\cos^2{\al\ov 2}+e^{-\eta}\sin^2{\al\ov 2}\right) \right\} 
\end{eqnarray}
with the abbreviation 
\begin{equation}
\gamma = e^{\eta}\sin^2{\al\ov 2}+e^{-\eta }\cos^2{\al\ov
2}.
\end{equation}
The diagonal elements are
\begin{equation}
\rho_0(x_{1},x_{1},\te) = \left({\Omega M\over \pi \hb
\gamma}\right)^{1/2} \exp\left\{-{\Omega M\over \hb\gamma}
x_{1}^{2}\right\}. 
\end{equation}

We can use~(\ref{NCDM}) to show that the relation
\begin{equation}
 \lb{TRHO} {\rm Tr}\left(\rho_{0}(\te)\right) = 1
\end{equation}
is satisfied as it should be for a normalized
state, where the notion $\rho_{0}(\te)=\rho_{0}(\vec{x},\vec{x}',\te)$
is used. The trace of $\rho_0^{2}(\te)$ can be obtained by
evaluating the integral
\begin{equation}\label{trace2}
{\rm Tr}\left(\rho_0^{2}(\te) \right) = \int
\rho_0(x_{1},x_{1}',\te) \rho_0(x_{1}',x_{1},\te)\;
dx'_{1}dx_{1}
\end{equation}
to end up with
\begin{equation}
\lb{TRSQ} {\rm Tr}\left(\rho_0^{2}(\te) \right) =
\left(1+ \sinh^{2}\eta \sin^{2}\alpha\right)^{-{1\ov 2}}.
\end{equation}
Clearly, the
traces are $\te$-independent and
therefore they are not affected by the non-commutativity
(\ref{nccoo}). The density matrix obtained by Kim {\em et al.} 
\cite{hkn99am} can be recovered by taking the limit $\te=0$.

\section{Wigner function for the NC system}

Next, we use the wave function~(\ref{NXWAV})
to determine the corresponding Wigner function,
which in general is defined by~\cite{wigner32}
\begin{equation}
\lb{WIN1} W_n(\vec{x}; \vec{p}) =  
\left({1\over \pi\hb}\right)^d \int
e^{-{2i\ov \hb}\vec{s}\cdot\vec{p}} 
\psi^{*}_n\left(\vec{x}-\vec{s}\right)
\psi_n\left(\vec{x} + \vec{s}\right)\; d^ds
\end{equation}
for the eigenfunctions  $\psi_n(\vec{x})$.
This definition is based on the operator formulation
of quantum mechanics. There is another definition
that can be used by introducing the $\hb$-star-product.
These two definitions are equivalent~\cite{group1}. One
concrete example was given by Dayi and Kelleyane~\cite{dayi} 
who evaluated the Wigner function
for an electron on the NC plane in the
presence of a magnetic field.

For~(\ref{NXWAV}) we evaluate the integral
\begin{equation}
\lb{NCWIN2} 
W_0(\vec{x}; \vec{p}, \te) =  \left({1\over \pi\hb}\right)^d
 \int e^{-{2i\ov \hb}\vec{s} \cdot \vec{p}} 
\psi^{*}_{0}\left(\vec{x}-\vec{s},\te\right) 
\psi_{0}\left(\vec{x} + \vec{s},\te\right)\; d^2s
\end{equation}
to get
\begin{equation}
\lb{wing}
W_0(\vec{x}; \vec{p},\te)= {1\over \pi} f_{0}(\vec{x},\te) 
g_{0}(\vec{p},\te) 
\end{equation}
where the functions $f_{0}(\vec{x},\te)$ and $g_{0}(\vec{p},\te)$ 
are
\begin{eqnarray}
f_{0}(\vec{x},\te)=\sqrt{M\Omega \ov \pi\hb} \exp\left\{-{M\Omega
\over \hb}\left[e^{\eta}\left(x_{1} \cos{\alpha\over 2} - x_{2}
\sin{\alpha\over 2}\right)^{2} +
e^{-\eta}\left(x_{1}\sin{\alpha\over 2} + x_{2}\cos{\alpha\over
2}\right)^{2} \right] \right\}
\nonumber \\
g_{0}(\vec{p},\te)=\sqrt{M\Omega \ov \pi\hb} \exp\left\{-{M\Omega
\over \hb}\left[e^{-\eta}\left(p_{1} \cos{\alpha\over 2} - p_{2}
\sin{\alpha\over 2}\right)^{2} +
e^{\eta}\left(p_{1}\sin{\alpha\over 2} + p_{2}\cos{\alpha\over
2}\right)^{2} \right] \right\}.
\end{eqnarray}
Integrating (\ref{wing})
over the variables $x_{2}$ and $p_{2}$, we obtain
\begin{eqnarray}
\lb{GSWF} 
\lefteqn{
W_{0}(x_{1},p_{1},\te) =\int W_{0}(\vec{x};
\vec{p},\te)\; dx_{2} dp_{2}
=
{M\Omega \over \pi\hb
\sqrt{1+ \sinh^{2}\eta \sin^{2}\alpha}} }
\nonumber \\
& &
~~~~~
\exp\left\{-\frac{M\Omega }
{\hb(\cosh\eta -\sinh\eta \cos\alpha)}x^{2}_{1}\right\}
\exp\left\{-\frac{M\Omega  } {\hb(\cosh\eta +
\sinh\eta \cos\alpha)}p^{2}_{1}\right\}
\end{eqnarray}
Of course at $\te=0$ we recover the standard Wigner function
for this system~\cite{hkn99am}.

One can use $W_{0}(x_{1},p_{1},\te)$ to verify
(\ref{TRHO}) and~(\ref{TRSQ}) by evaluating~\cite{knp91}
\begin{equation}\label{eq.23}
{\rm Tr}\left(\rho_0(\te)\right) = \int
W_0(\te)(x_{1},p_{1},\te) \; dx_{1} dp_{1} 
\end{equation}
as well as
\begin{equation}\label{eq.24}
{\rm Tr}\left(\rho_0^{2}(\te)\right) = 2 \pi \int
W_0^{2}(x_{1},p_{1},\te) \; dx_{1} dp_{1} .
\end{equation}
Indeed, these integrals confirm that the traces are
$\te$-independent.

\section{Uncertainty relation and entropy on $\RR_{\te}^2$}

The expectation value of any operator $A$ is
defined by the matrix element
\begin{equation}\label{aver00}
\langle A \rangle = \langle \psi| A |\psi \rangle
\end{equation}
where $|\psi \rangle $ is a normalized state. 
Also it can be expressed as the trace 
\begin{equation}\label{aver000}
\langle A \rangle = {\rm Tr}(A\rho)
\end{equation}
over the physical states.

The uncertainty of the operator $A$ is given by
\begin{equation}\label{uncer01}
\Delta A = \sqrt{\langle A^2 \rangle - \langle A \rangle^2}.
\end{equation}
Using either~(\ref{aver00}) or~(\ref{aver000}),
we calculate the product $\Delta x_1 \Delta p_1$ 
for the present system on $\RR_{\te}^2$
\begin{equation}
\label{uncer2}
{\Delta x_1 \Delta p_1 \ov \sqrt{1+ \sinh^{2}\eta
\sin^{2}\alpha }} =  {\hb\ov 2} \sqrt{1+\left({M\Omega\te\ov 2\hb}\right)^2}
= {\hb\ov 2} \sqrt{1+{a^2\ov 1 +a^2}}
 \end{equation}
where we have set $a={m\om\te\ov 2\hb}$. This relation  
can also be obtained by using the Wigner function~(\ref{GSWF}).

For $\te=0$, we recover the standard
result for the present system~\cite{hkn99am}
\begin{equation}\label{uncer22}
{\Delta x_1 \Delta p_1 \ov \sqrt{1+ \sinh^{2}\eta
\sin^{2}\alpha }} = { \hb\ov 2 }.
\end{equation}
For $\te\lga\infty$, we obtain 
\begin{equation}\label{uncer22in}
{\Delta x_1 \Delta p_1 \ov \sqrt{1+ \sinh^{2}\eta
\sin^{2}\alpha }} = { \hb\ov \sqrt{2} }.
\end{equation}
For small values of $\te$, (\ref{uncer2})
can be expanded as
\begin{equation}\label{uncer23}
{\Delta x_1 \Delta p_1 \ov \sqrt{1+ \sinh^{2}\eta
\sin^{2}\alpha }} \approx {\hb\ov 2}+ {\hb a^2\ov 4}.
\end{equation}
This shift can be
interpreted as a quantum correction to (\ref{uncer22}).
In Figure 1 we show how this relation depends on $\te$.

Finally, we calculate the entropy \cite{wiya63}
\begin{equation}
 S = - {\rm Tr} \left(\rho \ln\rho \right).
\end{equation}
for the NC system. Using the matrix elements
$(\rho_0(\te))_{mm}$ of $\rho_0(x_1,x_1',\te)$
we determine
\begin{equation}
S = - \sum_m (\rho_0{(\te)})_{mm} 
\ln(\rho_0{(\te)})_{mm}
= 2\left[\cosh^2\eta\ln(\cosh\eta) -
\sinh^2\eta\ln(\sinh\eta)\right]
\end{equation}
which is nothing but that obtained  by Kim {\it et
al.}~\cite{hkn99am}. Thus we conclude that $S$ is not 
changed by the non-commutativity (\ref{nccoo}).

\begin{figure}
\begin{center}
\includegraphics{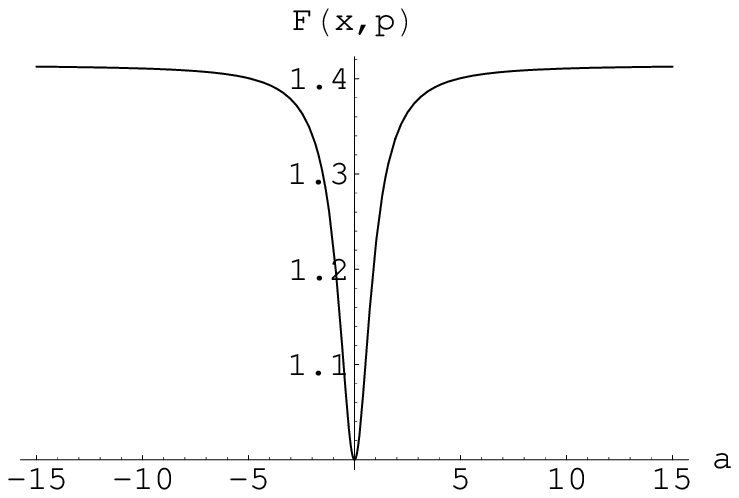}
\end{center}
\label{fig-nu-B} 
\caption{Variation of the uncertainty product  
${\mathrm{F(x,p)}}= {2\Delta
x_1 \Delta p_1 / \hb}$ in terms of $a= {m\om\te\ov 2\hb}$ for
$\eta=0$ or $\al=0$.}
\end{figure}

\section{ Particular values of $\al$}

To discuss some limits we distinguish two different
values of the mixing angle $\al$, i.e. $0$
and ${\pi\ov 2}$. In the former case we identify
the non-commutativity parameter $\te$ to
the squared magnetic length $l_B^2$. In this case 
we show that our system can be
linked to an electron on the plane in the
presence of the magnetic
field $B$.

\subsection{Case $\al={\pi\ov 2}$}

As we mentioned in our introduction, it is 
relevant to consider  
$\al={\pi\ov 2}$, because this is the case for
most physical systems described by two coupled harmonic
oscillators. In this limit, we get the ground state 
wave function 
\begin{equation}
\lb{PNXWAV} \psi_{0}(\vec{x},\te){\Big{|}}_{\al={\pi\ov 2}} =
\sqrt{M\Omega \ov \pi\hb} \exp\left\{-{M\Omega \over
2\hb}\left[(x_{1}^2+x_{2}^2)\cosh\eta
-2x_{1}x_{2}\sinh\eta\right] \right\},
\end{equation}
the density matrix
\begin{equation}
\rho_{0}(x_{1},x_{1}',\te){\Big{|}}_{\al={\pi\ov 2}}=
\left({M\Omega \over \pi
\hb \cosh\eta}\right)^{1/2} \exp\left\{{M\Omega \over
4\hb}\left[(x_{1} + x_{1}')^{2} \sinh\eta\tanh\eta +(x_{1}^2 +
x_{1}^{'2})\cosh\eta \right]\right\} 
\end{equation}
and the Wigner function
\begin{equation}
 W_{0}(\vec{x},\vec{p},\te){\Big{|}}_{\al={\pi\ov 2}} = 
{1\ov \pi} f_0(\vec{x},\te){\Big{|}}_{\al={\pi\ov
2}} g_{0}(\vec{p},\te){\Big{|}}_{\al={\pi\ov 2}}
\end{equation}
where $f_{0}(\vec{x},\te){\Big{|}}_{\al={\pi\ov 2}}$ 
and $g_{0}(\vec{p},\te){\Big{|}}_{\al={\pi\ov 2}}$ are
\begin{eqnarray}
f_{0}(\vec{x},\te){\Big{|}}_{\al={\pi\ov 2}} =
\sqrt{M\Omega \ov \pi\hb} \exp\left\{-{M\Omega \over
\hb}\left[(x_{1}^2+x_{2}^2)\cosh\eta - 2x_{1}x_{2}\sinh\eta\right]
\right\}
\nonumber \\
g_{0}(\vec{p},\te){\Big{|}}_{\al={\pi\ov 2}} =
\sqrt{M\Omega \ov \pi\hb} \exp\left\{-{M\Omega \over
\hb}\left[(p_{1}^2+p_{2}^2)\cosh\eta +2p_{1}p_{2}\sinh\eta\right]
\right\}.
\end{eqnarray}
The uncertainty relation simplifies to
\begin{equation}\label{uncer234}
{\Delta x_1 \Delta p_1 \ov \sqrt{1+ \sinh^{2}\eta
}}{\Bigg{|}}_{\al={\pi\ov 2}} = {\hb\ov 2} \sqrt{1+{a^2\ov 1 +a^2}}.
 \end{equation}

\subsection{Case $\al = 0$}

This case corresponds to the limit $c_3=0$. In the
standard geometry, the system becomes decoupled. However 
in the NC plane we still have an effective
coupling, which is $\te$-dependent.
Before going on, we note
that solving~(\ref{PARA}) we have two possibilities 
\begin{equation}\label{PARA1}
e^{\eta_1}= {c_1\ov \sqrt{c_1c_2}},
\qquad e^{\eta_2}= {c_2\ov \sqrt{c_1c_2}}.
\end{equation}
The expressions (\ref{PARA1}) can be simplified by linking
the present system to the Hall electron,
setting
\begin{equation}\label{fix}
\om=\sqrt{K\ov m}\ev {Be\ov 2mc}, \qquad \te_H= 4l_{B}^2
\end{equation}
where $l_{B}=\sqrt{\hb c\ov Be}$ is the magnetic length.
Then
\begin{equation}\label{PARA12}
c_1=c_2\ev \om\sqrt{m} , \qquad  e^{\eta_1}= e^{\eta_2}\ev 1.
\end{equation}
In this case, we recover the ground state wave function
of the Hall electron
\begin{equation}
\lb{PNXWAV2} \psi_{0}(\vec{x}, \te_H){\Big{|}}_{\al=0} =
\sqrt{m\om \ov \pi\hb} \exp\left\{-{m\omega \over
2\hb} (x_{1}^2+x_{2}^2) \right\}.
\end{equation}
This leads to the density matrix
\begin{equation}
\bra{l} \rho_{0}(x_{1},x_{1}', \te_H){\Big{|}}_{\al=0} 
= \left({m\om \over \pi
\hb }\right)^{1/2} \exp\left\{{m\omega \over 4\hb}
(x_{1}^2 + x_{1}^{'2}) \right\} .
\era
\end{equation}
The corresponding Wigner function reads 
\begin{equation}
 W_{0}(\vec{x},\vec{p}, \te_H){\Big{|}}_{\al=0} = 
{1\ov \pi}
f_{0}(\vec{x}, \te_H){\Big{|}}_{\al=0}
g_{0}(\vec{p}, \te_H){\Big{|}}_{\al=0}
\end{equation}
where $f_{0}(\vec{x}, \te_H){\Big{|}}_{\al=0}$ and
$g_{0}(\vec{p}, \te_H){\Big{|}}_{\al=0}$ are
\begin{eqnarray}
f_{0}(\vec{x}, \te_H){\Big{|}}_{\al=0} =
\sqrt{m\om \ov \pi\hb} \exp\left\{-{m\omega \over
\hb}(x_{1}^2+x_{2}^2) \right\}
\nonumber \\
g_{0}(\vec{p}, \te_H){\Big{|}}_{\al=0} =
\sqrt{m\om \ov \pi\hb} \exp\left\{-{m\omega \over
\hb}(p_{1}^2+p_{2}^2) \right\}.
\end{eqnarray}

\section{Conclusion}

We have investigated quantum mechanically a system of two
coupled harmonic oscillators on the non-commutative 
plane by requiring that 
the spatial coordinates do not commute 
and employing the star-product definition. 
By writing down its NC Hamiltonian
and making use of a 
suitable transformation, i.e.
a unitary rotation with the mixing angle $\al$, 
we have ended up with a diagonalized
system where the condition (\ref{acon}) for $\al$ was taken 
into account. 
By solving the eigenequations, the eigenstates and the
energy spectrum as well as the ground state wave function
are found to depend on the non-commutativity parameter
$\te$ and to coincide for $\te=0$ with those for the standard case. 
 This was used to determine 
explicitly the ground state wave function
of the NC system before the transformation, which
was the starting point of our interest. 

Subsequently, we have used the above tools to determine some
physical quantities corresponding to 
the ground state.
We have evaluated the corresponding
density matrix, which is $\te$-dependent.
However, we have shown that its traces, i.e. 
${\rm
Tr}\left(\rho_{0}(\te)\right)$ and ${\rm
Tr}\left(\rho_0^2(\te)\right)$, are not affected by 
the non-commutativity. Also the Wigner 
function was calculated for the NC system
and used to confirm the trace properties. We have
explicitly evaluated the uncertainty product and some
relevant limits were investigated. By taking small
values of $\te$, 
we have found that the standard relation was shifted by 
a term proportional to $\te^2$. This effect was
interpreted as quantum correction to the
normal case. Of course this shift vanishes for $\te=0$. 
We have found no $\te$-dependence for the entropy 
of the NC system, it is the same as
that for the normal case.

Finally, we have considered some limits of 
the parameters. In particular,
we have studied the case $\al={\pi \ov 2}$
which is relevant for many physical systems. 
In another interesting case, $\al={0}$,
the standard system becomes
decoupled. However, the NC system retains
an effective coupling due
to the parameter $\te$. By identifying $\te$ 
to the squared magnetic length, 
basic features of the Hall system 
were recovered, in particular its ground state 
and its Wigner function were given.

This work should be a good starting point to investigate
different issues related to the physical systems 
described by two coupled
harmonic oscillators. In particular, it can be used
to investigate different models proposed to study
many physical problems as mentioned 
in our introduction, see \cite{hkn89,kim89,hkn90}
and $[12-20]$, by deforming the spatial coordinates. 
We will return to these issues
and related matter in future.


\section*{{Acknowledgments}}
AJ's work is supported by Deutsche Forschungsgemeinschaft within
the Schwerpunkt ``Quantum-Hall-Effekt''. ELEH's work is supported
by AB-ICTP within the framework of the associateship scheme.

\end{document}